\documentclass[11pt]{article}
\usepackage{graphicx}
\usepackage{amsmath,amssymb}

\def\be{\begin{equation}}
\def\ee{\end{equation}}
\def\bes{\begin{eqnarray}}
\def\ees{\end{eqnarray}}
\def\tr{{\mathrm tr}}

\title{Quantum gravity in terms of topological observables}

\author{Laurent Freidel$^{1,2}$\thanks{email: lfreidel@perimeterinstitute.ca},
 Artem Starodubtsev$^1$\thanks{email: astarodubtsev@perimeterinstitute.ca} \\
\\ \centerline{\footnotesize \it $^1$Perimeter Institute for
Theoretical Physics} \\
\centerline{\footnotesize \it 35 King st N, Waterloo, ON, Canada
N2J 2W9. }\\
\centerline{\footnotesize \it $^2$Laboratoire de Physique, \'Ecole Normale Sup{\'e}rieure de Lyon} \\
\centerline{\footnotesize \it 46 all{\'e}e d'Italie, 69364 Lyon Cedex 07, France.}
}
\date{}

\begin{document}

\maketitle


\begin{abstract}
We recast the action principle of four dimensional General
Relativity so that it becomes amenable for perturbation theory
which doesn't break general covariance. The coupling constant
becomes dimensionless $(G_{Newton} \Lambda)$ and extremely small
$10^{-120}$. We give an expression for the generating functional of perturbation theory.
 We show that the partition function of quantum General
Relativity can be expressed as an expectation value of a certain topologically
invariant observable. This sets up a framework in which quantum gravity can be studied
perturbatively  using the techniques of topological quantum field theory.
\end{abstract}

\section{Introduction}
It has long been believed that quantum general relativity is
non-renormalizable. There is a solid argument for why it should be
so. The coupling constant of General Relativity, the Newton
constant, is dimensional and has to be multiplied by energy to
form the actual coupling. Therefore, at high energies the theory
becomes strongly coupled producing infinitely many types of
divergent Feynman diagrams. This is understood as the breakdown of
the theory below some scale sets by the Newton constant.

The above argument is not completely physical however. General
Relativity is the theory with no preexisting metric. And we can
ask: with respect to what metric we define the scale below which
the theory breaks down? The answer is that this is the metric
$g_{0ab}$ that we use as the background for the expansion and
defining the corrections to be quantized
\begin{equation}
g_{ab}=g_{0ab}+h_{ab}. \label{stupidformula}
\end{equation}
By fixing $g_{0ab}$  we fix not only a particular classical
solution, but also a coordinate system, and therefore break
general covariance. Thus, the argument for non-renormalisability
appears to rely on auxiliary structure which physics must be
independent of.

It is therefore worthwhile to explore the issue of
renormalizability from a point of view which doesn't use any
background metric. There are two possible outcomes of this effort.
Either we get an argument for non-renormalizability which would
use only relations between physical quantities, or we show that
non-renormalizability is an artifact of non-covariant quantization
based on (\ref{stupidformula}) and goes away in a background
independent quantization.

In a famous paper \cite{witten} Witten has shown that in 2+1
dimensional gravity, if we don't do the expansion
(\ref{stupidformula}), if we treat the whole geometry quantum
mechanically, thus keeping the theory general covariant, we can
avoid the ultraviolet problem. One can even show that the theory
exactly soluble.

And the natural question that arises, which is also addressed in
\cite{witten}, is whether we can do the same in 3+1 dimensions.
The immediate problem then is the following. If we look at the
action of 2+1 dimensional General Relativity in the triad-Palatini
representation
\begin{equation}
S=\int {\rm tr} (e \wedge dA + e\wedge A\wedge A)
\end{equation}
we see that the lowest order term in it is quadratic. Thus the
theory is nearly linear and we can apply standard quantum field
theory techniques to it. On the other hand the action of 3+1
dimensional General Relativity in the tetrad-Palatini
representation looks like
\begin{equation}
S=\int {\rm tr} (e\wedge e \wedge dA + e \wedge e\wedge A\wedge
A).
\end{equation}
The lowest order term in it is cubic. The standard quantum field
theory techniques are not applicable anymore. The only way out
seems to be to get a quadratic term in a action via the expansion
(\ref{stupidformula}), which leads to a non-renormalizable theory.
The conclusion of \cite{witten} is that 3+1 dimensional General
Relativity is non-renormalizable because it is too non-linear.

One of the questions that we address in this paper is: how
non-linear is 3+1 dimensional General Relativity?

Below, to avoid complications with using a non-compact group we
will consider Euclidian gravity with positive cosmological
constant.

\section{McDowell-Mansouri type BF-theory: How non-linear is 4 dimensional General Relativity?}

There are several formulations of 4 dimensional general relativity
known which do contain a quadratic term in the action \cite{plebanski}. They are
based on $BF$-theory plus a term which breaks topological
symmetry. A well known example is the so called  Plebanski action\footnote{A similar formulation of gravity works
in any dimension \cite{KPL}}:
\begin{equation}
S=\int (B^{\mu \nu}\wedge F_{\mu \nu}(\omega) +\phi^{\mu \nu
\alpha \beta} B_{\mu \nu} \wedge B_{\alpha \beta}) \label{pa}
\end{equation}
The first term in (\ref{pa}) is a $BF$-theory which is an exactly
soluble theory and one could think that we can use it as a free
field theory and treat the remaining term as a perturbation. There
is a problem with such a perturbation theory. This is the fact
that the second term actually imposes some constraints on $B^{\mu
\nu}$, as $\phi^{\mu \nu \alpha \beta}$ is a Lagrangian
multiplier. In a path integral this term becomes a delta function
of the constraints, and to treat it as a perturbation we would
have to expand a delta function in a power series around zero. But
such an expansion doesn't exist. What we would need for a
perturbation theory is an action principle in which General
Relativity would be represented as an exactly soluble theory plus
a regular interaction term. Such a formulation does exist and this
is the McDowell-Mansouri formulation of General Relativity
\cite{mm} rewritten as a $BF$-theory. Such kind of action
principle was also considered in \cite{phasetranse}.

Our starting point will be the $BF$-theory for $SO(5)$ group. Let
$T^{IJ}=-T^{JI}$ be ten generators of $so(5)$ Lie algebra, where
$I,J=1,...,5$ (see appendix \ref{so5con} for our conventions and physical
interpretation of other gauge groups).
The basic dynamical variables are
$so(5)$-connection $A^{IJ}$ and $so(5)$-valued 2-form field
$B^{IJ}$. The action principle is then
\begin{equation}
S=\int B^{IJ}\wedge F_{IJ}. \label{action0}
\end{equation}
Here $F_{IJ}=dA_{IJ}+A^{K}_{I}\wedge A_{KJ}$ is the $so(5)$
curvature.

The equations of motion following from the action (\ref{action0})
\begin{eqnarray}
F_{IJ}=0 \nonumber \\
d_A B_{IJ}=0 \label{eqs0}
\end{eqnarray}
mean that the connection $A^{IJ}$ is flat.

Now the statement is that if we break the $SO(5)$ symmetry in the
theory (\ref{action0}) down to $SO(4)$ we get the action of General
Relativity.

We add an extra term to the action (\ref{action0}) which depends
only on $B$-field and contains a fixed $SO(5)$ vector $v^A$
pointing in some preferred direction.
\begin{equation}
S_1=\int(B^{IJ}\wedge F_{IJ}-\frac{1}{2}B^{IJ}\wedge
B^{KL}\epsilon_{IJKLM}v^M). \label{action1}
\end{equation}
The $SO(5)$ symmetry is not a symmetry of the action
(\ref{action1}). It is broken down to $SO(4)$, the subgroup of
$SO(5)$ rotations which leaves $v^I$ immovable. For simplicity we
choose $v^I=(0,0,0,0,\alpha/2)$, where $\alpha$ is a fixed
dimensionless constant. The action (\ref{action1}) then becomes
\begin{equation}
S_1=\int (B^{IJ}\wedge F_{IJ}-\frac{\alpha}{4} B_{IJ}\wedge
B_{KL}\epsilon^{IJKL5}), \label{action11}
\end{equation}

To show that (\ref{action11}) is the action of General Relativity
we introduce the following notations for $4+1$-decomposition. Let
$i,j = 1,2,..4$ be four dimensional indices such that
$\epsilon^{ijkl}=\epsilon^{ijkl5}$. Then we can introduce
$so(4)$-connection $\omega^{ij}=A^{ij}$ and its curvature
$R^{ij}(\omega)=d\omega^{ij}+\omega^{i}_k \wedge \omega^{kj}$.
Also, we can introduce a frame field $e^i=l A^{i5}$, where $l$ is
a constant of the dimension of length, giving rise to a
four-dimensional metric $g_{\mu \nu}=e_\mu^i e_{\nu i}$.

In the above notations we have the following decomposition of
$so(5)$-curvature:
\begin{eqnarray}
F^{ij}(A)=R^{ij}(\omega)- \frac{1}{l^2}e^i \wedge e^j
\nonumber
\\
F^{i5}(A)=\frac{1}{l}d_\omega e^i \label{fdecomp}.
\end{eqnarray}
The equations of motion of (\ref{action11}) for $B^{5i}$
impose the torsion to vanish $d_\omega e^i =0$, this determines uniquely
the connection $\omega$ to be the spin connection.
Since the action is quadratic in the fields $B^{ij}$ we can
 solve the equations of motion for $ B^{ij}$
 and substituting them back into action, we
find
\begin{equation}
S_1=\frac{1}{4\alpha}\int F^{ij} \wedge F^{kl}
\epsilon_{ijkl}, \label{action21}
\end{equation}
where we used the notations introduced above. Finally, using
(\ref{fdecomp}) one can rewrite (\ref{action21}) as
\bes
S_1&=&\frac{1}{4\alpha}\int(R^{ij}-\frac{1}{l^2}e^i \wedge e^j)\wedge
(R^{kl}-\frac{1}{l^2}e^k \wedge e^l)\epsilon_{ijkl}
\nonumber \\
& =&
S_P + \frac{1}{4\alpha}\int R^{ij}(\omega) \wedge R^{kl}(\omega)\epsilon_{ijkl}
\label{action31}
\ees
Here
\begin{equation}
S_P=-\frac{1}{2G}\int( R^{ij}(\omega)\wedge e^k \wedge e^l
-\frac{\Lambda}{6} e^i \wedge e^j \wedge e^k \wedge e^l )
\epsilon_{ijkl} \label{actionp}
\end{equation}
is the Palatini action\footnote{
The normalizations are such that when written in the metric variables the Palatini
action is of the usual form
\be
S_P=-\frac{1}{G}\int \sqrt{g}(R -2\Lambda),
\ee
$R$ being the scalar curvature.}  of General Relativity
with nonzero cosmological constant. The role of the Newton
constant\footnote{ We work in
units where $c$ and  $16\pi\hbar =1$, so
$G$ means $ 16\pi G\hbar= l_p$ which is the Planck length.} is played by $G=\alpha l^2$
and the cosmological constant
is $\Lambda = 3/ l^2$. The constant $\alpha$ in (\ref{action0})
 is the square of the ratio of the Planck length over the cosmological radius,
$\alpha=G \Lambda/3 \sim 10^{-120}$ is dimensionless and extremely
small which makes it a good parameter for perturbative expansion.

The second term in the r.h.s. of (\ref{action31})
 is the integral of the Euler class. It is topological and its variation
 vanishes identically due to Bianchi identity.
 Thus the action (\ref{action1}) indeed describes General Relativity.

The main result of this section is that General Relativity in four
dimensions admits an action principle (\ref{action11}) which is
just barely non-linear, exactly as non-linear as that of 3
dimensional gravity. Also, (\ref{action11}) has a form of exactly
soluble theory plus a small correction. The correction is so small
that even if we neglect it we should give a good approximation to
the observed reality. And indeed we do, because the equations of
motion in this case are (\ref{eqs0}) which in the case of
$SO(4,1)$ gauge group have the only solution which is the deSitter
spacetime, which is very close to what we see.
This formulation of gravity is strikingly similar to a formulation of QCD in terms of the
Lagrangian $\tr(B\wedge F + g^2 B\wedge \star B)$. The difference coming from having
a quadratic form contracting the $B$ fields which is  strictly
positive and background dependent.

Despite the smallness of $\alpha$, however, there are many
situations in which the second term in the action (\ref{action11})
leads to  noticeable effects. This happens when some of the
components of $B$-field are large. Then the second term in
(\ref{action11}) which is quadratic in $B$ cannot be neglected as
compared to the first term which is linear in $B$, even though
multiplied by a tiny constant. In classical theory $B$ field
becomes large, for example, when we couple gravity to massive
matter sources \cite{classical}.

In quantum theory we have to take into account large fluctuations of $B$-field, thus
including the regime in which the theory becomes strongly coupled. This may lead to a
breakdown of perturbation theory. Most 'visible' is the contribution from the components
of $B$ which form the orbit of the 'translational' gauge group of the free field theory,
$B=d_A\phi$, which is broken by the interaction term.
To avoid this problem we need to find a way to suppress  large fluctuations
of $B$-field in a path integral. This can be done by a very natural modification of
the action principle considered in the
next section.


\subsection{Introducing the Immirzi Parameter}

In the previous section we have described gravity
in terms of a symmetry breaking perturbation of topological $BF$ theory.
In this section we generalize this construction to the case where the
topological field theory is $BF$ with a `cosmological term'.
As we will see this is necessary in order to regulate in a physical way
our perturbative expansion. We will also see that at the classical level
this allows us to introduce naturally an other dimensionless parameter which
appears in $4D$ gravity, the so called Barbero-Immirzi parameter.

The action principle for $SO(5)$ $BF$ theory with a cosmological term is \cite{Baez}
\begin{equation}
S=\int B^{IJ}\wedge F_{IJ} - \frac{\beta}{2} B^{IJ}\wedge B_{IJ}. \label{action4}
\end{equation}
The equations of motion following from this action  are
\begin{eqnarray}
F_{IJ}&=&\beta B_{IJ}, \nonumber \\
d_A B_{IJ}&=&0 \label{eqs01}.
\end{eqnarray}
Note that the first equation implies the second one.
This theory is invariant under local $SO(5)$ transformation, it is
topological due to the additional `translational' symmetry
labelled by a one form $\Phi^{IJ}$ valued in the Lie algebra\footnote{The non linear transformations
corresponding to this infinitesimal symmetry are given by
\bes
A&\rightarrow& A +\beta \phi,\\
 B&\rightarrow& B + d_A \phi + \frac{\beta}{2} {[}\phi,\phi{]}=
 B + \frac{F(A+\beta \phi) -F(A)}{\beta}.
 \ees
 \label{NLgauget}}
\bes
\delta A^{IJ}=\beta \Phi^{IJ},& &\quad \delta B^{IJ} =  d_A\Phi^{IJ}.
\ees
The gauge invariant observables of this theory are therefore gauge
invariant functions of $B^{IJ} - F^{IJ}/\beta$.
As before, we add an extra term to the action which breaks
the gauge symmetry down to $SO(4)$ gauge symmetry and also breaks
translational symmetry. Our proposal for a gravity action is
\begin{equation}
S_2=\int (B^{IJ}\wedge F_{IJ}- \frac{\beta}{2} B^{IJ}\wedge B_{IJ}
-\frac{\alpha}{4} B_{IJ}\wedge B_{KL}\epsilon^{IJKL5}) \label{action41}.
\end{equation}
 We can solve the equations of motion\footnote{We restrict to the case $\alpha^2\neq \beta^2$.
 Considering this case will lead to a self dual formulation of gravity.} for $ B^{IJ}$
 \bes
 B^{ij} &=& \frac{1}{\alpha^2 -\beta^2}(\frac{\alpha}{2} \epsilon^{ijkl}F_{kl} - \beta F^{ij}),\\
B^{5i} &=& \frac{1}{\beta } F^{5i},
\ees
and substitute them back into the action   (\ref{action41}), we get
\be\label{action42}
S_2=\int \left( \frac{\alpha}{4(\alpha^2-\beta^2)} F^{ij}\wedge F^{kl}\epsilon_{ijkl}
- \frac{\beta}{2(\alpha^2-\beta^2)} F^{ij}\wedge F_{ij}
+ \frac{1}{\beta} F^{5i}\wedge F_{5i}\right).
\ee
Using (\ref{fdecomp}) and introducing the Nieh-Yan class
$C= d_\omega e^i \wedge d_\omega e_i - R^{ij}\wedge e_i\wedge e_{j}$ \cite{NY},
we can rewrite this action in terms of gravity variables
\be\label{action43}
S_2=\widetilde{S}_P+\int \left(
\frac{\alpha}{4(\alpha^2-\beta^2)} R^{ij}(\omega) \wedge R^{kl}(\omega)\epsilon_{ijkl}
- \frac{\beta}{2(\alpha^2-\beta^2)}R^{ij}(\omega) \wedge R_{ij}(\omega)
+\frac{1}{\beta} C\right).
\ee
The last term is an integral of a linear combination
of the Euler class, the Pontryagin class and the Nieh-Yan class.
These are integer valued topological invariants with trivial
local variation.
The first term of action (\ref{action43})
\be
\widetilde{S}_P = -\frac{1}{2G}\int
\left(  R^{ij}(\omega)\wedge e^k \wedge e^l \epsilon_{ijkl} -
\frac{\Lambda}{6} e^i \wedge e^j \wedge e^k \wedge e^l \epsilon_{ijkl}
-\frac{2}{\gamma}R^{ij}(\omega)\wedge e_i \wedge e_j
\right) \label{actionfinal}
\end{equation}
is the  Cartan-Weyl action  of General Relativity
with nonzero cosmological constant and a nonzero Immirzi parameter
 $\gamma$, which is dimensionless \cite{Holst}.
The initial parameters $\alpha, \beta, l$ are related to
the physical parameters as follows
\be
\frac{1}{l^2}=\frac{\Lambda}{3},\qquad
\alpha = \frac{G\Lambda}{3(1-\gamma^2)},
\qquad \beta = \frac{\gamma G\Lambda}{3(1-\gamma^2)}.
\ee
Even if the term proportional to $\gamma$ is not topological (its
variation is non zero), it doesn't affect the classical equation of motion
when $\gamma^2 \neq 1$.
It plays no role in the classical theory of gravity
and it is therefore not constrained experimentally.
It is important to note however that this term, similarly to the theta
term in non abelian gauge theory, breaks CP symmetry.
Since this fact seems to have been unnoticed let us explain it in more
details. Suppose that we perform  an orientation
reversing diffeomorphism of our spacetime, lets call it a C-transformation.
All the terms in the action change signs since they are 4-forms, so C
is not a symmetry of our gravity action.
Lets now consider the discrete Lorentz transformations
$g_{ij}=\mathrm{diag}(-+++)$ or $g_{ij}=\mathrm{diag}(+---)$,
that we respectively called T or P transformation. The T
transformation changes only the sign of $e^{0}$ and $\omega^{0i}$
leaving all the other fields invariant. The first two terms in the
action change sign under P or T since they contain one epsilon tensor
contracting the Lorentz indices but the last term does not.
The action is not invariant under P or T but if we now consider
CP (or CT) we see that the first two terms in the action are left
invariant whereas the last one changes sign. In other word CP does not
affect $G$ or $\Lambda$ but changes the sign of the Immirzi parameter.
The CP symmetry is therefore realized only if $\gamma = 0 $ or
$\infty$.
When $\gamma=0$, which is the case studied previously,
we recover the case of usual metric gravity, where the torsion is identically $0$.
When $\gamma =\infty $ we recover the usual Cartan-Weyl gravity
where the torsion is free to fluctuate.
Any other value of $\gamma$ leads to a CP violation mechanism in quantum gravity
which is worth exploring.

Even if it doesn't affect the classical theory  the Immirzi parameter
deeply affects the quantum theory and labels inequivalent quantizations
in the context of kinematical loop quantum gravity.
Indeed it is known for a long time (see \cite{Ashtekar} for a review)
that this parameter modifies the symplectic structure
and this modification is not unitarily implementable at the quantum
level. It affects the prediction of the spectra of geometrical operators
and plays a key role in the black hole entropy calculation \cite{BH}.
This calculation suggests that $\gamma$ and $1-\gamma^2$ are of order unity.
One should keep in mind however, that the above conclusions are based on kinematical considerations,
i.e. before Hamiltonian constraint is applied.
And one open problem in this context is to understand wether the Immirzi
parameter really leads to inequivalent quantization once the
dynamics is fully taken into account or wether it can just be reabsorb into a redefinition
of the Newton constant. This point have already been raised at the kinematical level in
\cite{Alexandrov} where a seemingly more covariant approach to loop gravity leads to
a geometrical spectra independent of the Immirzi parameter.

A more direct way to understand why the Immirzi parameter should affects quantization is to
remark that $2/\gamma$ is proportional to the torsion square since
$\int R^{ij}\wedge e_i\wedge e_{j}=
\int d_\omega e^i \wedge d_\omega e_i$ up to a boundary term.
 $\gamma$ therefore controls the width of fluctuation of the
torsion at the quantum level. We have already remarked that if
$\gamma=0$, which is the case of metric gravity studied in the previous section,
 the torsion is not allowed to fluctuate.
 The mean value of the torsion is always
equal to $0$ irrespective of $\gamma$, this is why it doesn't
affect classical gravity. However a naive semiclassical
calculation shows that one should expect the two point function of
the torsion to be proportional to $\gamma$. Therefore $\gamma$
controls how strongly we suppress or not the torsion fluctuations
in the path integral.

In our context the Immirzi parameter appears to act as a physical regulator.
The role of the Immirzi parameter in this theory and its relevance to
the physical predictions will be explored in more detail in our next paper
\cite{pspinfoam}.

There is not yet any preferred experimental value for $\gamma$,
wether it is $0$, $\infty$ or the value suggested by loop quantum
gravity. Anyway, in all these cases\footnote{if $\gamma\rightarrow \infty$
both $\alpha$ and $\beta$ are sent to zero while the ratio
$\beta^2/\alpha$ tends to a finite value $G\Lambda/3$} $\alpha$
and $\beta$ are at most of the order $G\Lambda$ hence tiny
numbers.

\section{Formal setup for perturbation theory}
We first concentrate on the case $\beta=0$.
Let us rewrite the action (\ref{action11}) in an index free
form
\begin{equation}
S_{GR}=\int {\rm tr}(B\wedge F(A)-\frac{\alpha}{4} B \wedge B \gamma_5).
\label{action12}
\end{equation}
Here $B=B^{IJ}T_{IJ}/2$ and $A=A^{IJ} T_{IJ}/2$, where
\begin{equation}
T_{IJ}=\frac{1}{4}[\gamma_I,\gamma_J] \label{generators}
\end{equation}
are $so(5)$-generators in the fundamental representation and
$\gamma_I$ are $\gamma$-matrices satisfying $\{\gamma_I, \gamma_J
\}=2\delta_{IJ}$. The insertion of $\gamma_5$ in the second term
of (\ref{action12}) breaks $SO(5)$ symmetry down to $SO(4)$.

We will be calculating the path integral for the action
(\ref{action12})
\begin{equation}
Z_{GR} = \int {\cal D} A {\cal D} B \exp( S_{GR}) \label{zgr}.
\end{equation}
 Following \cite{genfunct} we will treat the $BF$
term in (\ref{action12}) as free field theory and the second term
as a perturbation. Define the generating functional which is the
path integral for the $BF$ theory with an extrinsic source as
\begin{equation}
Z(J) =  \int {\cal D} A {\cal D} B \exp \Big( \int {\rm
tr}(B\wedge F(A)- B \wedge J)  \Big), \label{gf}
\end{equation}
where $J$ is an $so(5)$-valued 2-form field. Then the path
integral for General Relativity can be obtained by including the
interaction by differentiating with respect to the sources.
\begin{eqnarray}
Z_{GR}=\exp \Big( \int {\rm tr}(\frac{\alpha}{4} \frac{\delta}{\delta
J}\wedge \frac{\delta}{\delta J} \gamma_5) \Big) Z(J)
\bigg\vert_{J=0}\label{zgr1}.
\end{eqnarray}
The perturbation theory can be obtained by expanding the exponent
in (\ref{zgr1}) in a power series
\begin{eqnarray}
Z_{GR}=\sum\limits_{n} \frac{1}{n!}\Big( \int {\rm tr}(\frac{\alpha}{4}
\frac{\delta}{\delta J}\wedge \frac{\delta}{\delta J} \gamma_5)
\Big)^n Z(J) \bigg\vert_{J=0} \label{zgr2}.
\end{eqnarray}
As $\alpha \ll 1$ we expect the sum to be dominated by the lowest order
terms.

\subsection{Computing the generating functional}\label{gfcomp}
We know show that the generating BF functional can be exactly
evaluated. We start with
\begin{equation}
Z(J) =  \int {\cal D} A {\cal D} B \exp \Big( i\int {\rm
tr}(B\wedge F(A)-\frac{\beta}{2}B \wedge B - B \wedge J)  \Big).
\label{gfb}
\end{equation}
The action is quadratic in the $B$ field we can integrate it out
by replacing it by its classical value
\be
\beta B^{IJ} = F^{IJ}(A) -J^{IJ},
\ee
the action becomes
\be\label{ActJ}
S_{J}= \frac{1}{2\beta} \int {\rm
tr}\big((F(A)-J)\wedge (F(A)- J)  \Big).
\ee
Its equations of motion are
\be\label{eomj}
d_{A}J=0,
\ee
and we denote by ${\cal M}_{J}$ the solution space.
In order to solve these equations lets introduce
a linear operator mapping Lie algebra valued 1-forms to Lie algebra
valued three forms
\bes
L_{J}: \Omega_{1}({\cal G})&\rightarrow & \Omega_{3}({\cal G})\\
 A & \rightarrow & {[}J,A{]}.
\ees
The space of three forms $L_{\mu\nu\rho}\in\Omega_{3}({\cal G}) $
is isomorphic to the space of densitized vector
$\tilde{L}^\alpha = 1/2 \epsilon^{\alpha \mu\nu\rho}L_{\mu\nu\rho}$.
$L_{J}$ is a square matrix whose matrix elements can be explicitly
written as
\be
\tilde{L}_{J}^{(\nu AB)}\big(^\mu_{CD}\big)= \epsilon^{\alpha\beta \mu \nu}
J_{\alpha \beta {[}C}^{\!
{[}A}\delta^{B{]}}_{D{]}}.
\ee
For a generic $J$, we expect $L_{J}$ to be  invertible,
in this case there is a unique connection solution of \ref{eomj}
\be
a_{J}= L_{J}^{-1}(dJ)
\ee¥
We expand $A= a_{J}+ a$ and the action around this solution
\bes\label{actionJ}
2\beta S_{J}= \int_{M}¥ {\rm
tr}\big((F(a_{J})-J)\wedge (F(a_{J})- J) +
2(d_{a_{J}}a +\frac{1}{2}{[}a,a{]})\wedge (F(a_{J})- J) \\
+(d_{a_{J}}a +\frac{1}{2}{[}a,a{]})\wedge(d_{a_{J}}a +\frac{1}{2}{[}a,a{]}).
\Big)
\ees
This expansion can be drastically simplified.
First we can integrate by part the second term in the action
using the equation of motion and the Bianchi identity
$d_{a_{J}}J= d_{a_{J}}F(a_{J})=0$.
We can also integrate by part the third term in the action
by introducing the Chern-Simons functional
\be
CS_{J}(a) = {\rm tr}\big(a\wedge d_{a_{J}}a +\frac{1}{3} a\wedge{[} a{,}a{]} \big),
\ee
its derivative is given by
\be
dCS_{J}(a) = {\rm tr}\big(
(d_{a_{J}}a +\frac{1}{2}{[}a,a{]})\wedge(d_{a_{J}}a +\frac{1}{2}{[}a,a{]})
+ {[}a,a{]}\wedge F(a_{J})
\big).
\ee
The action (\ref{actionJ}) can then be written as a sum of a boundary term
\be
\int_{\partial M} CS_{J}(a) + 2 {\rm tr}\big(a\wedge (F(a_{J})-J)\big)
\ee
and a bulk action which remarkably is quadratic
\be\label{quadaction}
2\beta S_{J}= \int_{M} {\rm
tr}\big((F(a_{J})-J)\wedge (F(a_{J})- J) +
 a\wedge{[}J,a{]} \big).
 \ee
We can then get an exact expression for the generating functional
\be
Z(J) = \frac{\exp \Big( \frac{i}{2\beta}\int_{M}{\rm tr}
(F(a_{J})-J)\wedge (F(a_{J})- J)
\Big)}{\sqrt{\mathrm{det} L_{J}}}.
\ee
In the denominator we have the determinant of the operator $L_J(x,y)\equiv L_J(x)\delta(x,y)$.

If $J$ is such that $L_J$ is not invertible we can still carry out the computation.
In this case ${\cal M}_J$, the space of solutions of (\ref{eomj}), is an affine space
of non zero dimension, its tangent space is the kernel of $L_J$.
The action (\ref{ActJ}) now possess an extra gauge invariance
\be
\delta A = a,\quad  \mathrm{with} \, \, a\in \mathrm{ker}({L_J}).
\ee
 We denote by $a_J$ any solution of (\ref{eomj}),
 and expand as before $A=a_J +a$, we still get the quadratic action (\ref{quadaction}).
 The integration over $a$ now gives
 \be
Z(J) =\int_{{\cal{M}}_J/G_J} da_J \frac{\exp \Big( \frac{i}{2\beta}\int_{M}{\rm tr}
(F(a_{J})-J)\wedge (F(a_{J})- J)
\Big)}{\sqrt{\mathrm{det}' L_{J}}}.
\ee
where $\mathrm{det}'$ denotes the determinant of $L_J$ acting on a orthogonal subspace of
 $Ker(L_J)$, $G_J=\{g/ gJg^{-1}=J\}$ is the subgroup preserving  $J$,
 and ${\cal{M}}_J/G_J$ is the space of solution modulo  gauge transformation.

\section{Topological effective action}

We now want to study further the path integral and the effect of
the gauge symmetry breaking term. We  suppose in the following
that $\beta =0$. The gauge parameters are pairs  $g,\phi$ were
$g\in SO(5)$ and $\phi$ is a one form Lie algebra valued. The BF
action $S_{BF}(B,A)$ is invariant under the transformation \bes
A&\rightarrow& ^gA= g A g^{-1} +gdg^{-1}, \\
B&\rightarrow& g(B-d_{A}\phi)g^{-1}. \ees We can split the
integration over $A,B$ into an integration over the gauge
equivalence class $[A],[B]$ of the BF symmetry and an integration
over the gauge  parameters $g,\phi$.

 The integration measure decomposes, by the standard Faddev-Popov argument as $ {\cal D} A
{\cal D} B = {\cal D} [A] {\cal D} [B] {\cal D} g {\cal D} \phi$
and the path integral becomes
\begin{equation}
Z =  \int {\cal D} [A] {\cal D} [B] e^{ S_{BF}(B,A) + s(A,B)} .
\label{gaugeaction}
\end{equation}
where $s(A,B)$ is an effective action obtained by integration over
the gauge degree of freedom, explicitly
 \be\label{effaction}
e^{is(A,B)} =  \int {\cal D} g {\cal D} \phi e^{ -\frac{\alpha}{4}
\int d^{4}x¥{\rm tr}( |B-d_{A}\phi|\gamma(g))}
\ee
Where we define
$ \gamma(g) \equiv g^{-1}\gamma_{5}g$ which is a unit
vector\footnote{$\gamma_{5}$ is left invariant by an $SO(4)$
subgroup so $\gamma(g)$ determines a point in $S^{4}=
SO(5)/SO(4)$} in $\mathbb{R}^{5}$ and for every $SO(5)$ valued 2
form $B$ we define the vector density $|B|(x)=
|B|_{M}(x)\gamma^{M}$ by\footnote{If we spell out the indices this reads
\be |B|= \gamma^{M}\epsilon_{IJKLM} \epsilon^{\mu \nu \rho \sigma}
B_{\mu\nu}^{IJ} B_{\rho\sigma}^{KL} \ee }
\be {\rm tr}(B\wedge B \gamma_{M}) =
|B|_{M}d^{4}x
\ee

In  order to understand the form of
this effective action we will  look at the partial effective
action obtained by integrating only $g$ and only $\phi$.
As we will see each partial integration
is one loop exact, which means that it localizes on its classical
solutions and that its evaluation is given by its stationary
phase evaluation (see \cite{witten2} for a discussion of
localization in QFT). This is clear for the integration over
$\phi$ since the action is quadratic in $\phi$ but it also happens
for the integration over $g$.

\subsection{spontaneous symmetry breaking}
The action $\int{\rm tr}( |B-d_{A}\phi|\gamma(g))$ is ultralocal
for $g$, it doesn't contain any derivative  acting on $g$ and we
can therefore  understand the localization property of the path
integral by looking at the final dimensional analog
\be
\int_{SO(4)} dg e^{i{\rm tr}( |B| \gamma(g))} =
\frac{e^{i||B||}}{2||B||} - \frac{e^{-i||B||}}{2||B||}.
\ee
The RHS
of this expression is  the semi classical evaluation of the
integral. This is clear since the equation of motion of ${\rm tr}(
|B| \gamma(g))$ gives $[\gamma(g), |B|]= 0$. The solutions when
$||B||^{2}\equiv |B|_{M}|B|^M \neq 0$ are given by
\be \gamma(g)=
\pm \frac{|B|}{||B||}.
\ee
The action evaluated on this solution
is $\pm ||B||$ which reproduces the terms in the exponential, the
denominator comes from the evaluation of the quadratic
fluctuations around the solution.

Similarly, the equations of motion of the continuum action
$\int{\rm tr}(|B|\gamma(g))$ (supposing $\phi=0$ for simplicity)
 are $[\gamma(g), |B|]= 0$.
 The solutions when
$||B||^{2}\equiv |B|_{M}|B|^M \neq 0$ are given by \be
\gamma(g)(x)= \pm \frac{|B|(x)}{||B||(x)}. \ee The sign is a
priori $x$ dependent, but if we restrict to continuous solution
for $g$ we have only two solutions\footnote{this problem disappear
when we consider negative cosmological constant, in this case
$\gamma(g)$ belongs to an hyperboloid and there is only one
solution to the equation
$[\gamma(g), |B|]= 0$.}.

The localization property of the integral therefore suggest that
the gravity effective action obtained by integration over $g$ is
(if one keep one branch)
\begin{equation}
S_{GR}=\int {\rm tr}(B\wedge F(A))-\frac{\alpha}{4} ||B||(x)d^4x.
\end{equation}
This action is now $SO(5)$ invariant whereas gravity is only
$SO(4)$ invariant, it sounds therefore strange at first that we can
recover gravity from this action. In order to understand this let
us first check that the equation of motion for this action are
equivalent to Einstein's equation.

The action is defined for all $B$, however it is differentiable only when
$||B||(x)\neq 0$ which we now suppose holds true. The equations of motion are given by
\bes
F^{IJ} &=&\frac{\alpha}{2}\epsilon^{IJKLM}B_{KL}n_M, \\
d_AB^{IJ}&=&0.
\ees
where we denote $ n_M \equiv\frac{|B|_M}{||B||}$.
Given this unit vector we define
\bes\label{var1}
{e^I} &\equiv  d_A n^I = dn^I + A^{IJ}n_J,
&\omega^{IJ} \equiv  A^{IJ} + n^Ie^J-n^Je^I,\\
&b^I\equiv  B^{IJ}n_J,
&b^{IJ} \equiv  B^{IJ} + n^Ib^J-n^Jb^I.\label{var2}
\ees
We denote $F^{IJ}$ (respectively  $R^{IJ}$) the curvature of the connection
$A$ (respectively $\omega$).
We have the following identities
\bes
F^{IJ}n_J = d_A e^I = {d_{\omega}e^I}, \\
R^{IJ}- {2} e^{[I}\wedge e^{J]}= F^{IJ}-2d_{\omega}e^{[I}n^{J]},
\ees
the bracket denotes antisymetrisation.
From these identities it is clear that $R^{IJ}n_J=0$, also $d_\omega n_I=0$, so $\omega$ is a $SO(4)$ connection
preserving the direction $n^I$.
In terms of the variables (\ref{var1}, \ref{var2})  the equations of motion read
\bes
d_{\omega}e^I&=&0, \label{1}\\
 \frac{1}{2\alpha}\epsilon_{IJKLM}(R^{IJ}- {2} e^{I}\wedge e^{J})n^M &=&  b_{KL}, \label{2}\\
d_AB^{IJ}n_J= d_\omega b^I -b^{IJ}\wedge e_J&=&0,\label{3}\\
d_{\omega} b^{IJ}+2 e^{[I}\wedge b^{J]} &=&0.\label{4}
\ees
The first equation tells us that $\omega$ is the spin connection if
 the frame field $e$ is invertible.
If one take the derivative $d_\omega$ of (\ref{2}) one obtain that
$d_\omega b_{KL}=0$ since $d_\omega R^{IJ} =0 $ by Bianchi identity, $d_\omega e^I=0$
by the torsion free equation and $d_\omega n_I=0$ by construction.
The equation (\ref{4}) then imply that $b^I=0$ when $e^I$ is invertible.
This means that $b^{IJ}\wedge e_J=0$ by equation (\ref{3}) which is
equivalent, due to (\ref{2}), to the Einstein equation
\be
\epsilon_{ijkl} (R^{ij}- {2}e^{i}\wedge e^{l})\wedge e^k =0,
\ee
where the indices $i,j,k$ label vectors orthogonal to $n$.

One sees that the equation of motion of the $SO(5)$ invariant theory are equivalent to Einstein
equation when $e$ is invertible. Even if the action is invariant under $SO(5)$ gauge symmetry
the solutions of this action spontaneously breaks this symmetry by choosing a preferred direction
in the internal space proportional to $|B|$.
The same results apply for $\beta\neq 0$.

 On shell
we have that $||B||d^4x= \frac{1}{\alpha^2}|\epsilon_{ijkl}F^{ij}\wedge F^{kl}| $.
Any $SO(4)$ bivector $B^{ij}$ can be decomposed into a self dual and anti self dual
part $B=B_+ +B_-$, using this decomposition for spatial and internal indices of
$F_{\mu \nu}^{ij}= $ we can write decompose $F$ as $F= W_+ +W_-+ \phi + \phi_0$
where $W_-^{ij}$ is a symmetric traceless tensor labelling the $5$ self dual components of the Weyl tensor,
$\phi^{ij}$ a traceless tensor labelling the trace free part of the Ricci tensor
and $\phi_0$ is the scalar curvature.
In term of these components we have
$||B||=4!det(e)/\alpha^2( (W_+)^2+(W_-)^2 +(\phi_0)^2-(\phi)^2)$
The components $\phi,\phi_0$ are zero by the Einstein equation, thus
$||B||$ is zero if and only if The Weyl tensor vanish that is only
if $F=0$ and our spacetime is spherical\footnote{in the Lorentzian case the condition is less restrictive since
$W_{\pm}$ are complex conjugate} .
We therefore see that the presence of a spontaneous symmetry breaking is
equivalent in the Euclidean case to the existence of a non trivial gravitational field.

\subsection{Gravity as a non local topological theory}

We now consider the construction of the effective action coming from the integration of the
translational symmetry parameter for $g$ fixed. We discuss  the case $\beta=0$.
\be\label{transint}
e^{is(A,B)=}  \int {\cal D} \phi e^{ i\frac{\alpha}{4}
\int d^{4}x {\rm tr}( |B-d_{A}\phi|\gamma_5)}.
\ee
This integral being quadratic the integral localizes on the classical solution if any.
The equation of motion are given by
\be
d_A\{B,\gamma_5\}=\Delta_A \phi,
\ee
where $\Delta_A$ is the differential operator
$\Delta_A \phi = d_A\{d_A\phi,\gamma_5\}$.
If one uses the $4+1$ decomposition $A^{IJ}=(\omega^{ij},e^i)$ and $
\phi^{IJ}=(\phi^{ij},\phi^i)$ we can write these equation in components.
\bes
d_\omega B^{ij} = d_w (d_\omega\phi^{ij}-2e^{[i} \wedge \phi^{j]}) \\
\epsilon_{ijkl}B^{ij}\wedge e^k= \epsilon_{ijkl}(d_\omega\phi^{ij}-2e^{i}\wedge \phi^{j})\wedge e^k.
\ees
If $\Delta_A$ is invertible we can uniquely solve this equation.
Lets denote  $\varphi\equiv \Delta_A^{-1}(d_A\{B,\gamma_5\})$  a solution of these equations
and define
\bes
\bar{B} = B -d_A\varphi,
\ees
by construction $\bar{B}$ is a solution of $ d_A\{\bar{B},\gamma_5\}=0$
and if we insert the previous decomposition in the integral (\ref{transint}) we
can factorize $\bar{B}$ out of the integral
\be
e^{is(A,B)}= e^{i \frac{\alpha}{4} \int {\rm tr}(\bar{B}\wedge \bar{B}\gamma_5) }
\int {\cal D} \phi e^{ -\frac{\alpha}{4}
\int d^{4}x {\rm tr}( |d_{A}\phi|\gamma_5)} =
 \frac{e^{i\frac{\alpha}{4} \int {\rm tr}(\bar{B}\wedge \bar{B}\gamma_5) }}{\sqrt{{\rm det}(\Delta_A})}
\ee
where $\Delta_A$ is the differential operator
$\Delta_A \phi = d_A\{d_A\phi,\gamma_5\}$.

One sees that the integration over the gauge modes produces for us an effective action
\be\label{effactionfin}
\tilde{s}(A,B)=\frac{\alpha}{4} \int {\rm tr}(\bar{B}\wedge \bar{B}\gamma_5) =
\frac{\alpha}{4} \int {\rm tr}\Big(({B}\wedge {B}\gamma_5)+ \frac{1}{2}(d_A\{B,\gamma_5\}\Delta_A^{-1}d_A\{B,\gamma_5\})\Big).
\ee
This action is invariant under the translational gauge symmetry of $BF$ theory
which is the symmetry that makes $BF$ theory topological but by construction its
partition function is the one of gravity.
$BF$ theory does not carry local degrees of freedom whereas gravity does,
the catch is that the effective action
is a non local observable for $BF$ theory since it involves the propagator of $\Delta_A$.
It is important to remark that this action is still quadratic in $B$ since $\Delta_A$ is a linear operator.
It is not clear however, wether we can explicitely do the $g$ integration of the action (\ref{effactionfin}).

In the derivation of the effective action we have assumed that $\Delta_A$ is an invertible operator
that is that there is no non trivial solution to the equation
\be
d_A \{ d_A\phi,\gamma_5\}=0.
\ee
We
expect it to be true for a generic choice of $A$ (as long as $e=d_A\gamma_5$ is invertible).
We are now  going to give an argument in favor of this claim, keeping in mind that
it will be interesting to have a proper characterization of the connections for which it holds.

Before doing so, let us first study a particular case  where on the contrary $\Delta_A$
is not invertible.
 We  will now show that if $A$ is a flat $SO(5)$, then the gravitational waves
 around this connection are in one to one correspondence
with the kernel of $\Delta_A$.
If we start from a Cartan Weyl formulation of gravity (\ref{actionp}), with $\Lambda=3$,
the equation of motions are
\be
d_\omega(R^{ij}-e^{[i}\wedge e^{j]})=0, \quad
\epsilon_{ijkl}(R^{ij}-e^i\wedge e^j)\wedge e^k=0.
\ee
 In the first equation which comes from variation with respect to $\omega^{ij}$ we have added for convenience a term
 trivial by Bianchi identity.
 These equations can be written in a compact form
 \be
 d_A\{F(A),\gamma_5\}=0,
 \ee
by using the notation of eq.(\ref{fdecomp}).
Given a gravity solution $A=(\omega^{ij},e^i)$ we can look for `graviton solution', i-e
infinitesimal perturbation $\delta A $ such that $A+\delta A $ is a solution of Einstein equations
to first order.
The equation for the perturbation is
\be
\Delta_A \delta A = [\{F(A),\gamma_5\}, \delta A].
\ee
Therefore, if the original space time is a four sphere: $F(A)=0$, and $\delta A$ is a
graviton solution, then $\phi= \delta A$ is in the kernel of $\Delta_A$.
Even in Euclidean space where there is no graviton $\Delta_A$ is not invertible
around a flat $SO(5)$ connection since infinitesimal diffeomorphism $\delta A = {\cal L}_\xi A$ are in the
kernel of $\Delta_A$.
Away from a spherical space this is no longer true:
the graviton Laplacian is now
$\tilde{\Delta}_A = \Delta_A + [\{F(A),\gamma_5\}, \cdot]$, infinitesimal diffeomorphisms
$\delta A = {\cal L}_\xi A$ are in the kernel of $\tilde{\Delta}_A$ but not of $\Delta_A$.
This can be easily understood from the fact that the action
$S(A,\phi)=\int {\rm tr}(d_A\phi^{ij} \wedge d_A\phi^{kl})\epsilon_{ijkl}$
is not invariant under diffeomorphism or $SO(4)$ gauge transformation acting on $\phi$ alone unless
$A$ is chosen to be fixed by a combination of diffeomorphism and gauge transformation.
This is the case for a flat connection since we have in this case
${\cal L}_\xi A - d_A (i_\xi A) =i_\xi(F(A))=0$, with $\xi$ a four vector and $i_\xi $ denotes the interior
product\footnote{ In the case of Flat $SO(5)$ connection this action is clearly
invariant under the transformation $\delta\phi =d_A \psi$}.
In general  $\Delta_A$ being  non invertible means that $S(A,\phi)$
possess some gauge invariance. We expect all possible gauge invariance of such an action  to come
from a restriction of the diffeomorphism group times the local rotation group.
Only some special connection will have such a subgroup of invariance like the flat connection but also
some special holonomy connections. For a generic $A$ there is no such invariance and
therefore we expect $\Delta_A$ to be invertible.

\subsection{Towards background independent perturbation theory}
In the previous section we have shown that the partition function of quantum General Relativity
can be expressed as an expectation value of a topologically invariant observable
given by (\ref{effactionfin}). The insertion of such an observable into topological field theory
can be understood as introducing an infinite-dimensional moduli space labelled by $\phi$ and $\gamma(g)$ and
then integrating over it. This moduli space contains all the physical degrees of freedom of four dimensional
General Relativity. Looking at the effective action (\ref{effactionfin}) one can see that unlike in ordinary
quantum field theory the gravitational degrees of freedom are generically non-local and do not allow  to
form  point-like excitation. The non-locality of fundamental gravitational excitation was argued on a
different basis in
\cite{topexc}. This is a realization of equivalence principle in quantum gravity
saying that a local free falling observer can never see gravitational effects.

For practical calculations we need a perturbation theory which would decompose the infinite dimensional moduli
space defined  into a set of finite dimensional moduli spaces in each of
which the integration can be explicitly performed.
This requires the exchange of the expansion of the interaction term in a power
series and integration over the broken gauge degrees of freedom.
Such an exchange is valid if all the resulting integrals converge.
The later is not always the case.
This can be seen if $\beta=0$, in this case we can perform directly the integral
over $\phi$. If we do the perturbative expansion before integrating these degrees of freedom we get meaningless
expression since the integral over $\phi$ diverges for large values of $\phi$.

The story is however quite different if $\beta\neq 0$.
First, at some naive level we see that in this case the large fluctuation of $B$-field
are suppressed due to strong oscillatory behavior of the integral by the quadratic term proportional to
$\beta$ which is a non positive but definite quadratic form.
This strongly suggest that Immirzi parameter act as a physical regulating parameter of
our perturbation theory.

One can also understand this by looking at the action for the $BF$ gauge degrees of freedom
which is, in the case $\beta\neq 0$, given by (see footnote \ref{NLgauget})
\be
\frac{\alpha}{4}\int d^4x \tr(|B-\frac{F(A)}{\beta} + \frac{F(A+\beta \phi)}{\beta}|\gamma(g)).
\ee
In this case the action is non quadratic in $\phi$ and we cannot hope to solve it directly.
The only possibility is then to construct perturbatively the partition function and
 do the perturbative expansion before the integration.
One clearly sees from this expression that integration over  $\beta \phi$
should now be understood as a integration over a $SO(5)$ connection.
In order to do the integration we need to specify the measure we
 use to integrate over the space of gauge invariant connection\footnote{the action is gauge invariant
 after integration over $g$}.
 We need a measure which is diffeomorphism invariant and normalized, such a measure has
been proven (under additional technical hypothesis) to exist and to be unique for a compact
gauge group \cite{Lost}: this is the Ashtekar-Lewandowski measure.
This suggests that the techniques of loop quantum gravity and spin foam model are adapted \cite{genfunct}
to describe our perturbative expansion and lead to a finite result.

In section \ref{gfcomp} we have calculated the generating functional needed for such a perturbation theory.
One can notice that the resulting perturbative expansion we are proposing is  non standard,
 since the insertion of a non zero $J$ breaks a gauge symmetry which is restored in the limit
  $J\rightarrow 0$. At each order in perturbation theory the derivatives over $J$ insert an operator
  which is slightly breaking the original $BF$ gauge symmetry.

At zero-order the result of the computation is simply given by $Z(0) =\int_{{\cal{A}}/G} dA$
which is the integral
over all gauge connection. If one uses the Ashtekar-Lewandowski measure this gives
 $Z(0)=1$, which is the correct result for trivial topology.
An other way to get the same result is to consider that this integral should be gauged
  fixed by the topological $BF$ symmetry and as such it is just equal to $1$.
At each order of the perturbation theory one would then compute the expectation value in $BF$ theory of a
operator which is of polynomial order in the $B$ fields and gauge fixed the residual $BF$ symmetry.
The gauge fixing procedure of $BF$ is now well understood in the spin foam context \cite{diffeo}.
This shows once again,that the techniques of  spin foam models are well
adapted \cite{genfunct} to describe our perturbative expansion.
At each order of perturbation theory a triangulation can be chosen to do the explicit computation, and the
topological symmetry which is acting away from the source should insure that the result is independent
of the triangulation.
This is contrast with the usual spin foam model which usually depends on the chosen triangulation,
and one has to average over triangulations using the
perturbative expansion of an auxiliary Field theory \cite{deprov}.

We can also understand in this context why the case $\beta\neq 0$  is much more regular.
It has been conjectured for a long time that the computation of the
partition function of $BF$ model with non zero `cosmological term' is realized by a state sum model
build on a quantum group with $q =\exp(i\beta)$ roots of unity. This has been proven recently by Barrett et al
\cite{Barrettetal} for the case of the group $SU(2)$, presumably this results holds for any
compact group.
Therefore the sums entering the computation of the BF observables in the spin foam context
are all expected to be finite.

Our next paper \cite{pspinfoam} is devoted to study in more details the perturbation theory in the context of
spin foam.

{\bf Acknowledgments:}
We would like to thank Lee Smolin for support and discussions.

\appendix

\section{SO(5) conventions}\label{so5con}
$T_{IJ}=-T_{JI} $ with $I=1,\cdots,5$ are the ten generators of $so(5)$. They satisfy
the algebra
 \be
 {[}T_{IJ},T_{KL}{]}= \eta_{JK} T_{IL} -\eta_{IK}T_{JL}+\eta_{IL}T_{JK} -\eta_{JL}T_{IK},
 \ee
 where $\eta_{IJ} =\delta_{IJ}$ in the case of $so(5)$.
 The corresponding theory of gravity is Euclidean with a positive cosmological constant,
 i-e `spherical gravity'. This is the one we focus on in the main text.
 If we want to describe Lorentzian gravity and/or  other sign of cosmological constant
 one should consider metric of different signatures, the cases of interest for gravity are:
 $SO(4,1)$, where $\eta=diag(++++-)$ which describes Euclidean gravity with a negative cosmological constant
 (i-e `hyperbolic gravity').
$SO(1,4)$ where $\eta= diag(-++++)$ which describes Lorentzian gravity with a positive cosmological constant,
i-e `de Sitter gravity'.
$SO(3,2)$ where $\eta= diag(-+++-)$ which describes Lorentzian gravity with a negative cosmological constant,
i-e `AdS gravity'.
One can split the generators of in terms of $so(4)$ generators
$T_{ij}$, $i=1,\cdots,4$ and `translation' generators
$ P_i = T_{i5}/l$, where $l$ is a length scale (cosmological length scale in our context).
The algebra reads
\bes
 {[}T_{ij},T_{kl}{]}&=& \eta_{jk} T_{il} +\cdots,\\
  {[}T_{ij},P_{k}{]}&=& \eta_{ik} P_{j} -\eta_{jk} P_{i},\\
   {[}P_{i},P_{j}{]}&=& -\frac{\eta_{55}}{l^2} T_{ij}.
 \ees
The $so(5)$ can be represented in terms of $\gamma$ matrix
\be
T_{IJ}=\frac{1}{4}{[}\gamma_I,\gamma_J{]}
\end{equation}
where
$\gamma_I$ are $\gamma$-matrices satisfying $\{\gamma_I, \gamma_J\}=2\eta_{IJ}$.

\end{document}